\documentclass[fleqn,10pt]{wlscirep}
\usepackage[utf8]{inputenc}
\usepackage[T1]{fontenc}
\usepackage{graphicx}
\usepackage{subcaption}
\usepackage[linesnumbered, boxed, resetcount]{algorithm2e}
\title{Epidemics on the Move: How Public Transport Demand and Capacity Shape Disease Spread}

\author[1,2]{László Hajdu}
\author[2]{Jovan Pavlović}
\author[1,2]{Miklós Krész}
\author[3,*]{András Bóta}
\affil[1]{Innorenew CoE, Livade 6a Izola, SI-6310, Slovenia}
\affil[2]{University of Primorska, FAMNIT, Glagolja\v{s}ka 8, SI-6000 Koper, Slovenia}

\affil[3]{Luleå University of Technology, 97187 Luleå, Sweden}

\affil[*]{andras.bota@ltu.se}


\begin{abstract}
Understanding the dynamics of passenger interactions and their epidemiological impact throughout public transportation systems is crucial for both service efficiency and public health. High passenger density and close physical proximity has been shown to accelerate the spread of infectious diseases. During the COVID-19 pandemic, many public transportation companies took measures to slow down and minimize disease spreading. One of these measures was introducing spacing and capacity constraints to public transit vehicles. Our objective is to explore the effects of demand changes and transportation measures from an epidemiological point of view, offering alternative measures to public transportation companies to keep the system alive while minimizing the epidemiological risk as much as possible.  
\end{abstract}
\begin{document}

\flushbottom
\maketitle

\thispagestyle{empty}

\section*{Introduction and Background}

Recent years have highlighted the importance of understanding the epidemiological transmission dynamics in public spaces, especially within public transportation systems. Due to the mass of interacting and co-travelling passengers, these environments become potential accelerators and hotspots for the spreading of different viruses, especially in the case of a global pandemic. During the COVID-19 outbreak, physical distancing became one of the most commonly used non-pharmaceutical measures to prevent the spread of the disease \cite{tirachini2020covid}. A good overview of COVID-19-related restrictions can be found here \cite{GKIOTSALITIS2021374}. In addition to causing efficiency and operational problems, these measures usually redirect passengers towards other travel options or block them from being able to travel in the city, causing disruptions and potentially a greater epidemiological risk in other environments. Therefore, exploring and understanding how the combination of fluctuations in passenger behavior and restrictive measures affect the spread of diseases is vital to providing an innovative and safe solution for society.

Agent-based models give us reliable solutions to approximate the behavior of public transportation users in the city. The structure of the public transportation system, including timetables, vehicle trips, or stations, defines a unique restricted environment where the agents/passengers are using the services, moving between stations, and interacting with each other on vehicles. \cite{Kerr} On the other side of the system is the demand, which defines how and when the passengers would like to move within the city, satisfying their objectives. Public transportation systems, however, have many parameters, resulting in a quite compound, sensitive, and hard-to-predict complex system especially if we aim to find epidemiological effects of those parameters. In the case of a pandemic, there can be changes in many parts of the system, including both structure (due to the restrictions) and demand (due to changed passenger needs) \cite{apple_mobility_2020, google_mobility_2020}. Therefore, we are aiming to simulate different scenarios in both structure and demand to understand the complex behavior of the system as a whole.

The range of different epidemiological models is quite broad, from compartmental models \cite{Brauer2008} to Independent Cascade\cite{kempe} or Linear Threshold \cite{granovetter} models. Usually, compartmental models, more precisely, SI, SIR, or SEIR models, are used to simulate the spread of disease, making them the most accepted approach to approximate outbreak scenarios. Originally, compartmental models were introduced to simulate the spread of a disease in a homogeneous space using differential equations; however, later, the concept was extended to networks \cite{walter2012compartmental, kempe, Bóta_Krész_Pluhár_2013}. Many existing studies also use agent-based simulations combined with compartmental models to simulate the spread of infectious diseases and information on contact networks, helping to predict and evaluate intervention strategies \cite{coleman1996medical, hasan2011contagion}. Smart card data \cite{sun2014efficient} and activity-based travel models \cite{Lam2003, Roorda2009} capture detailed passenger contact patterns, essential for developing epidemic surveillance and containment strategies. Lastly, simulating epidemiological models on contact networks requires transmission risk values to be available between the nodes of the network. Multiple models have been proposed to estimate these values based on real-life observations \cite{Sun2013, Bta2011SystematicLO,6932999}. 

However, creating accurate real-world contact networks from this data poses challenges, such as computational complexity and privacy issues \cite{cattuto2010dynamics, christakis2010social}. Indeed, there are also approaches to address privacy issues in collecting data and tracking infection and contacts in these data sources \cite{2020_Buchanan}. Nevertheless, it can be stated that this is a critical part of the data collection if we aim to create a real-like simulation environment. A good overview of the different methods and challenges of collecting human mobility data can be found here \cite{collecting}. A more precise interaction network allows us to model the epidemiological spread more realistically \cite{MO2021102893, Arenas2020.03.21.20040022, PARE2020345}. The literature offers a wide range of overview papers connected to this area\cite{10.3389/fpubh.2024.1367324, LIU2022100030, PUJANTEOTALORA2023104422}; however, neither of them focuses on the effect of different restrictions or measures. In summary, most approaches focus on analyzing historical use cases together with already executed restriction scenarios \cite{histo1, histo2,Murano2021}.

If we exclude the historical analyses, the literature could be better when it comes to simulating and evaluating the effects of different measures on an existing public transportation system from the epidemiological point of view. One study proposed an agent-based solution for exploring the impact of these measures but on a higher level, simulating the mobility patterns of agents in Italy \cite{FAZIO2022101373}. The paper \cite{LUO2022103592} introduces a framework to optimize public transit flow during pandemics, balancing mobility and safety by controlling transit routes. Naturally, there are other approaches trying to target similar questions \cite{10.1371/journal.pone.0260919,KAMGA202125}. However, literature lacks approaches predicting and evaluating the epidemiological spread in case of different hypothetical scenarios on an existing system. Overview papers related to this topic can be found in the following articles \cite{Ayouni2021, Konstantinos}.

In this paper, continuing the path of our previous research \cite{bota2017b_modeling, hajdu2020discovering}, we use simulation models to create different scenarios adjusting transit demand and interventions across the San Francisco Bay Area's public transport system to explore the effect of changes on the infection rate and the efficiency of the network. Using these simulations, we construct temporal contact networks representing potential passenger interactions specific to each scenario. We first run epidemiological simulations on these networks. Then we evaluate how the different demand- and intervention-related changes affect the spread of the viruses and the efficiency of the system. Finally, we try to find the balance between preventative measures and maintaining operational efficiency in public transportation systems. In the following section, we introduce the travel demand model, the structure of the contact network, and the epidemiological modelling algorithm. 

\section*{Methods}

In this paper, we use activity-based travel demand models to simulate highly probable passenger paths on the public transport network of San Francisco Bay Area, using real-life data. These models were complemented by schedule-based transit assignment models that provide accurate travel time estimates and transfer waiting times. After simulating the behavior of the passengers in the city, we constructed contact networks, computed properties of these networks, and analyzed the service efficiency of the system in each proposed scenario. Outbreak processes were then simulated using the discrete SIR model to assess infection spread and identify the effects of demand changes and restriction-related scenarios. The next section introduces the travel demand model we used to simulate passenger behavior.

\subsection*{Travel Demand Model}

We used the FAST-TrIPs \cite{Khani2013} route choice model during our work. This model assigns individual passengers to their final path choices using hyperpath probabilities and simulates them using a mesoscopic transit passenger simulation module. The model captures time-dependent daily service variability and focuses on specific transit vehicle trips to reflect real-world passenger route choices based on the timetable and the schedule of the service. Due to the unavailability of a calibrated transit route choice model, we used the model from \cite{Khani2014}. The route choice utility function is the following:

$$u = t_{IV} + 1.77t_{WT} + 3.93t_{WK} + 47.73X_{TR}$$
where $u, t_{IV}, t_{WT}, t_{WK}$ and $X_{TR}$ represent path utility, in-vehicle time, waiting time, walking time, and number of transfers in a transit path, respectively. To understand the underlying algorithm, FAST-TrIPs operates on a network of nodes representing stops. Each vehicle trip belonging to a specific transit route is associated with the stops it serves, along with the corresponding arrival and departure times. Each stop contains information about the vehicle trips serving it. Additionally, transfer links connect stops where passengers can switch vehicles. This structure allows for precise modeling of vehicle movements and passenger transfers across the transit network.

At the core of FAST-TrIPs is the Transit Hyperpath Algorithm, which constructs a subnetwork of probable transit routes and assigns probabilities to these routes using a logit route choice model. The algorithm calculates hyperpaths by considering user-preferred arrival times and waiting time windows, enabling the simulation of passenger journeys focusing on real-time decision-making and path selection. Passenger movements are then modeled using a pre-estimated route choice model incorporating in-vehicle time, waiting time, walking time (for access and transfers), and transfer penalties.

The model output contains the individual passenger trajectories, including their walking from the source to the bus stop, boarding and changing vehicles, walking between vehicle trips, and the final walk to the destination. The simulation represents the passenger movements on the public transportation system for one full day. The next section introduces the structure of the contact network focusing on the connection between the output of the FAST-TrIPs model and the network structure.

\subsection*{Contact Network}

As discussed in the previous section, the output of the FAST-TrIPs model describes the trajectories together with the exact vehicle trips, walks, and timestamps for each passenger using the transit system. The output can be further processed, creating a network structure we call the contact network. The objective of this network is to describe passenger interactions on vehicle trips. More precisely, let $G = (V,E)$ be an undirected network where each passenger using the public transportation system is considered a node; therefore, $V$ equals the set of traveling passengers, and $E$ is the set of co-traveling among passenger pairs which means that edges connect any two passengers who share a vehicle trip for a positive time. Let $VT$ be the set of vehicle trips. The name vehicle trip refers to a specific route with a particular departure time and is unique to a single vehicle. Additionally, let $e_p$ be the vehicle trip that connects two passengers, $e^{t_s}_{j}$ and $e^{t_e}_{v}$ the starting and ending time of the common part of their travel on vehicle trip $j$ where $\forall e \in E$ and $\forall p \in VT$. Using $e^{t_s}_{j}$ and $e^{t_e}_{v}$, we were also able to calculate the elapsed time the passengers were in contact on the actual vehicle trip. Therefore, edges also describe the strength of the connection between passengers, which can later be used to define a probability that expresses the likelihood of passing an infection through the connection. In summary, throughout the research, the contact network is used to define the fine structure of pairwise passenger connections.

\begin{figure}[!ht]
\centering
\includegraphics[width=0.6\linewidth]{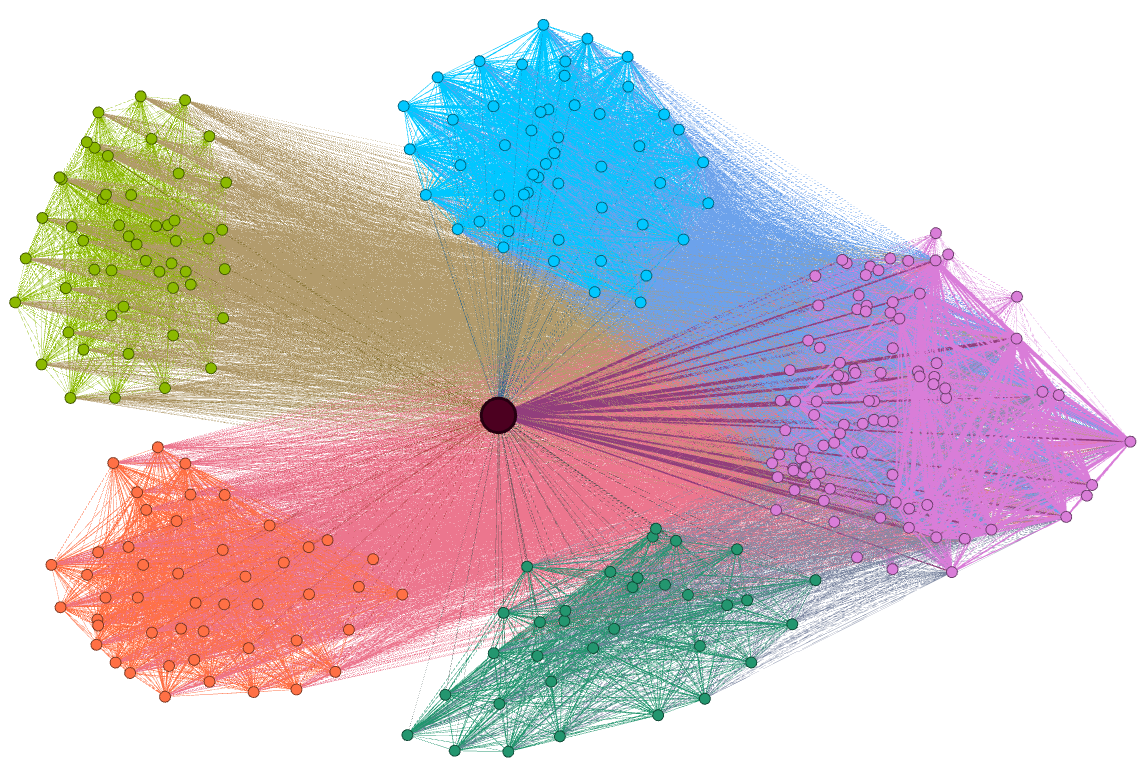}
\caption{Neighbourhood of the highest degree passenger in the contact network. Each node represents a passenger, while each edge is a connection between them on a vehicle trip they used together.}
\label{fig:highpass}
\end{figure}

Figure \ref{fig:highpass} shows a subgraph of the contact network around the highest degree passenger. Different colors correspond to different vehicle trips. It can be seen that the passenger with the highest degree shares many different vehicle trips with different strongly connected groups on the transit system. In the following section, we introduce the epidemiological model we used to simulate the spread of a virus in a case of an outbreak using the contact network.

\subsection*{Epidemiological Modeling}

We use the susceptible-infected-removed (SIR) compartmental model to represent the different epidemiological scenarios and changes in the transit system. Before we apply the SIR model to the contact network $G(V, E)$, we assign a transmission probability \( w_e \in [0,1]\) to each edge $e \in E$, representing the likelihood of infection spreading between connected nodes. According the SIR model definition, nodes in the network can exist in one of the three states: susceptible (S), infected (I), or removed (R). Nodes also have behaviors: a susceptible node can become infected; a removed node was infected once but is no longer contagious, while the infected state corresponds to the case when the node is currently spreading the infection to its neighbors.

At the beginning of each simulation, we choose an initial seed set $A_0$ that defines the group of initially infected passengers. The initial seed set can be selected through preliminary knowledge, simulating a historical epidemiological scenario, knowing the origin vehicle trip, and following the steps of the infection through the different parts of the system. On the other hand, the initial seed set can also be chosen randomly, focusing on the structural behavior of the network and introducing an additional stochastic parameter to the modeling. In this study, we selected the initial seed set randomly to explore the structural properties of the public transit system.

During each simulation iteration, infected nodes attempt to transmit the infection to susceptible neighbors based on the edge transmission probability $w_e$. If the transmission is successful, the susceptible neighbor transitions to the infected state in the following iteration, continuing the spread of the infection. The infected nodes usually remain in this state for $ \tau_i $ iterations before transitioning to the removed state.

The model behaves stochastically, meaning outcomes vary across simulations. By running the simulation $k$ times, we can estimate each node's probability of becoming infected by calculating the proportion of simulations in which the node becomes infected. 

In this application, each iteration corresponds to a single day, and nodes are assumed to remain infected for five days. As a result, we did not transition passengers to the removed state throughout our simulations. In this way, since each passenger gets the infection only once, our model is identical to the SI variant of the SIR model. Algorithm~\ref{algo:sir_simulation} summarizes the simulation process. In the following section, we introduce the calculation of the edge probabilities in the network.

\begin{algorithm}
    \small
    \SetAlgoLined
    \KwData{Contact network \( G \), edge infection probabilities \( w_e \), initial infected count \( n \), time steps \( T \), infection period \( \tau_i \)}
    \KwResult{1-0 array of length \(|V(G)|\) indicating which node was infected by the end of the simulation.}
    Initialize counters for each node with value \(-1\)\;
    Select \( n \) random nodes as initially infected\;
    Initialize set \( \textit{initvert} \) containing selected nodes\;
    Set \( \text{current\_time} \) to 0\;
    \While{\(\textit{initvert}\) is not empty \textbf{and} \( \text{current\_time} < T \)}{
        Initialize new set \( \text{newset} \) as empty\;
        \ForEach{node in \( \text{initvert} \)}{
            Get list of neighbors of node\;
            \ForEach{neighbor in neighbors}{
                \If{counter[neighbor] < 0 \textbf{and} random() \( \leq w_e[(node, neighbor)] \)}{
                    Increment counter[neighbor]\;
                    Add neighbor to \( \text{newset} \)\;
                }
            }
            Increment counter[node]\;
            \If{counter[node] \( \leq \tau_{i} \)}{
                Add node to \( \text{newset} \)\;
            }
        }
        Set \( \textit{initvert} \) to \( \text{newset} \)\;
        Increment \( \text{current\_time} \) by 1\;
    }
    Create array \( \textit{results} \) where each element is 1 if counter[node] \( \geq 0 \), otherwise 0\;
    \caption{SIR Model Simulation}
    \label{algo:sir_simulation}
\end{algorithm}

In order to assign realistic transmission probabilities to edges in the contact network, we use the function introduced in \cite{Sun2013}. We assign probabilities to edges in the contact network using the following formula:

    $$w_e = \min(P_{max}, \frac{P_{max}}{D_{max}} \cdot D_e)$$

Where \(P_{max}\) is the maximum possible transmission probability, \(D_{max}\) represents the duration threshold after which an edge is assigned the maximum probability, and \(D_e\) is the contact duration associated with edge \(e\). In the next section, we introduce the Input data and the different reduced demand and vehicle capacity scenarios.

\section*{Input data}

The travel demand model works with multiple input files that define the public transportation system and demand. The General Transit Feed Specification (GTFS and GTFS PLUS)\cite{GTFS2024} is a standardized format for defining transit schedules and geographical information. On the other hand, the transit demand data contains information about the trips individual passengers want to make throughout the day, including trip origins, destinations, and preferred arrival times. Additionally, path weights associated with in-vehicle time, waiting time, walking time, and transfer penalties must be specified as input. In summary, GTFS PLUS defines the public transportation network of the city, while the demand defines the different travel needs of the passengers.

In our use case, we used the General Transit Feed Specification of the San Francisco Bay Area in California from 2017. The transportation system includes 854 routes for bus, heavy and light rail, and ferry traffic. Altogether, 36,058 trips serve 6,181 stops over a 24-hour weekday. Since we could not access real-world demands, we used data generated for 2017 by the SF-CHAMP travel forecasting tool \cite{SFCTA2024}. As path weights, we used data from the previous study \cite{Khani2014}, corresponding to the Austin, Texas region. The following figure \ref{fig:pubnetwork} shows the public transit system, including the routes and the stops in the San Francisco Bay Area, California.

\begin{figure}[ht]
\centering
\includegraphics[width=0.7\linewidth]{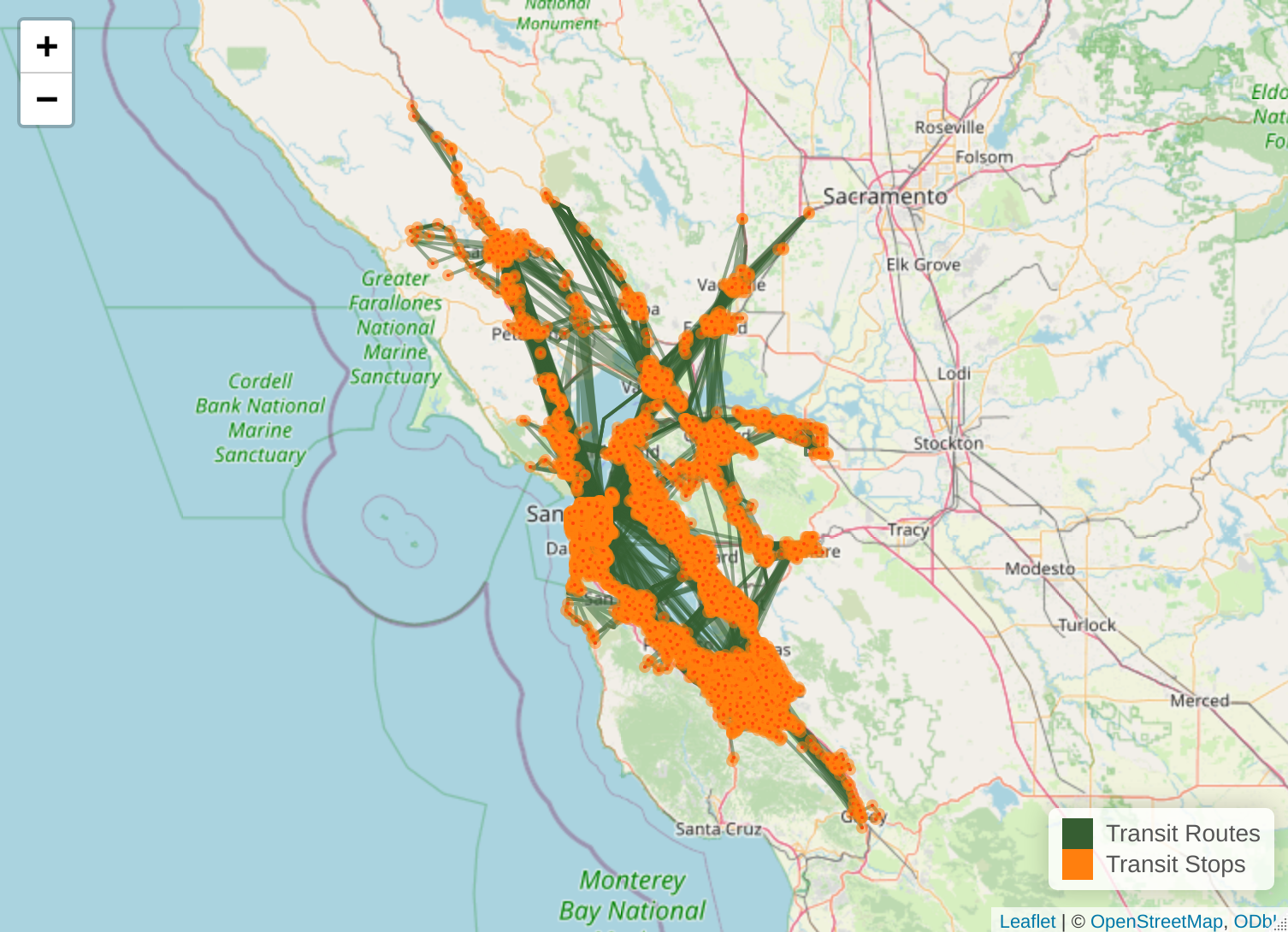}
\caption{Visualization of transit system}
\label{fig:pubnetwork}
\end{figure}

\begin{figure}[!ht]
\centering
    % First row: Two figures
    \begin{minipage}[b]{0.30\linewidth}
        \centering
        \includegraphics[width=\linewidth]{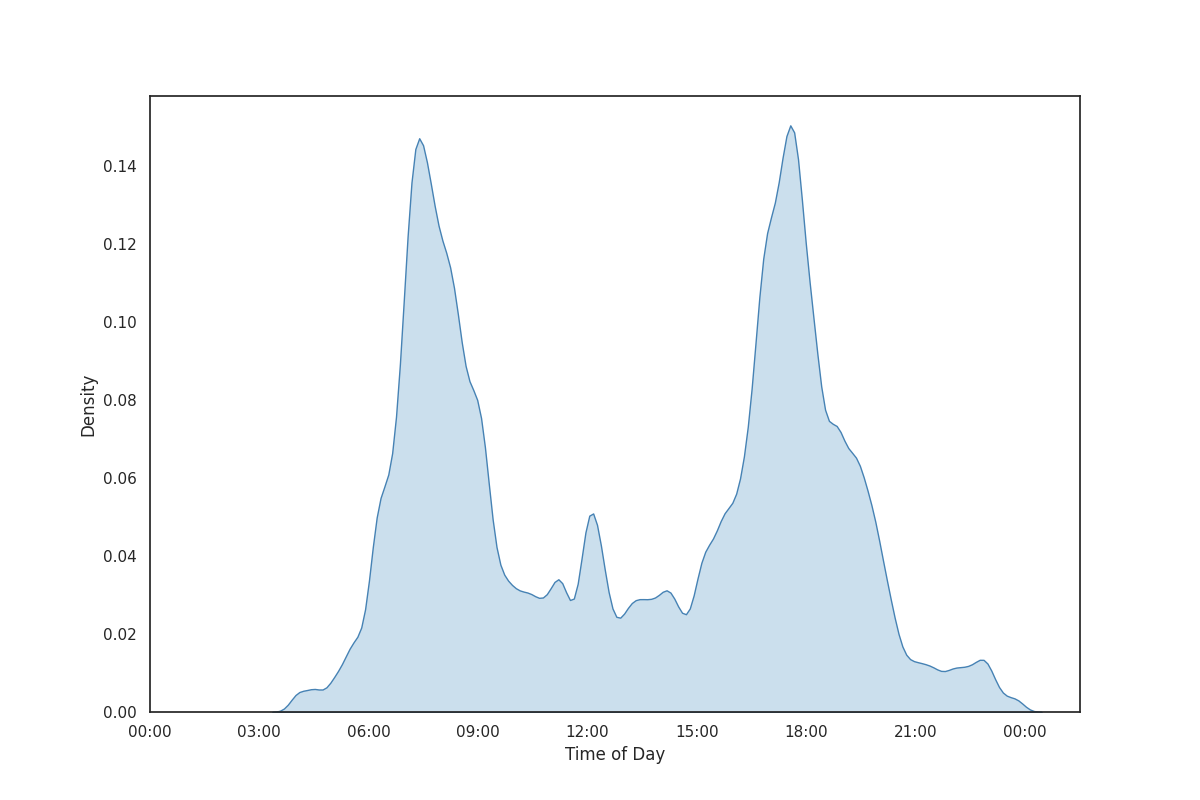}
        \subcaption{Distribution of contact stat times}
        \label{cntct}
    \end{minipage}
    \hspace{0.01\linewidth}
    \begin{minipage}[b]{0.30\linewidth}
        \centering
        \includegraphics[width=\linewidth]{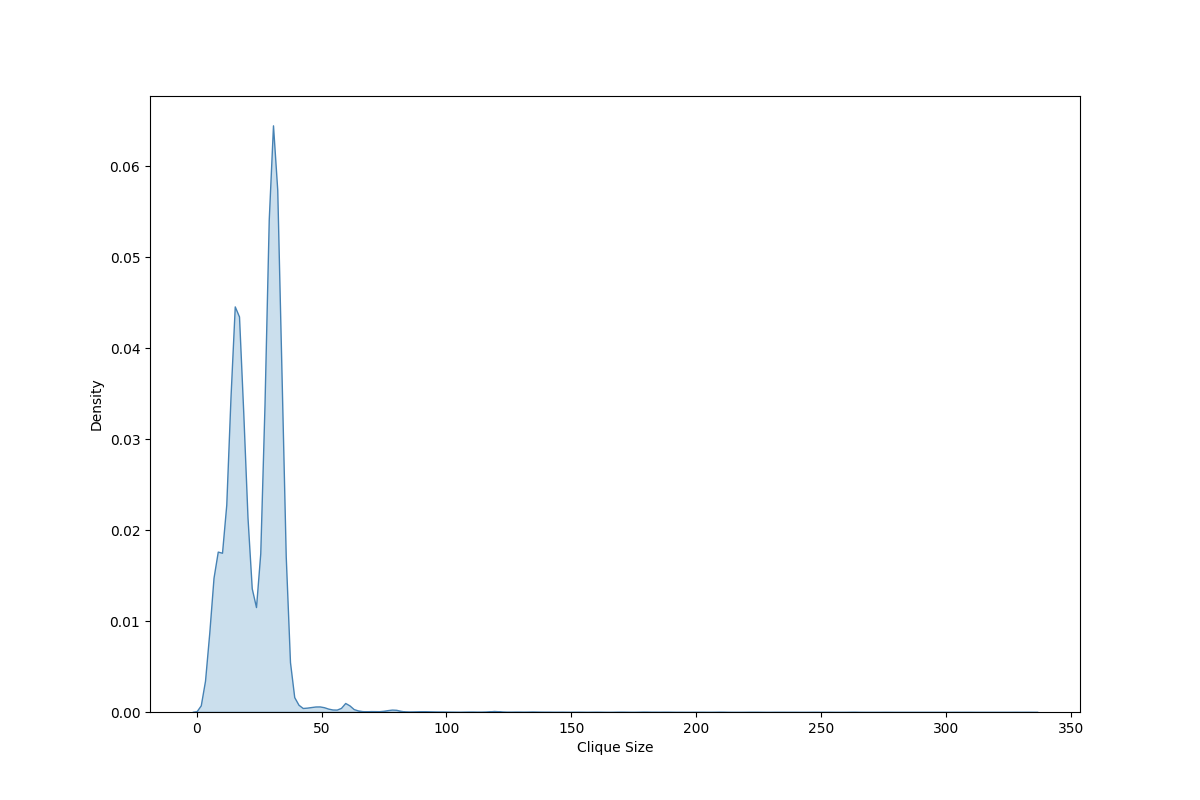}
        \subcaption{Distribution of clique sizes}
        \label{clique}
    \end{minipage}
    
   \vspace{0.01\linewidth}
   
    \begin{minipage}[b]{0.30\linewidth}
        \centering
        \includegraphics[width=\linewidth]{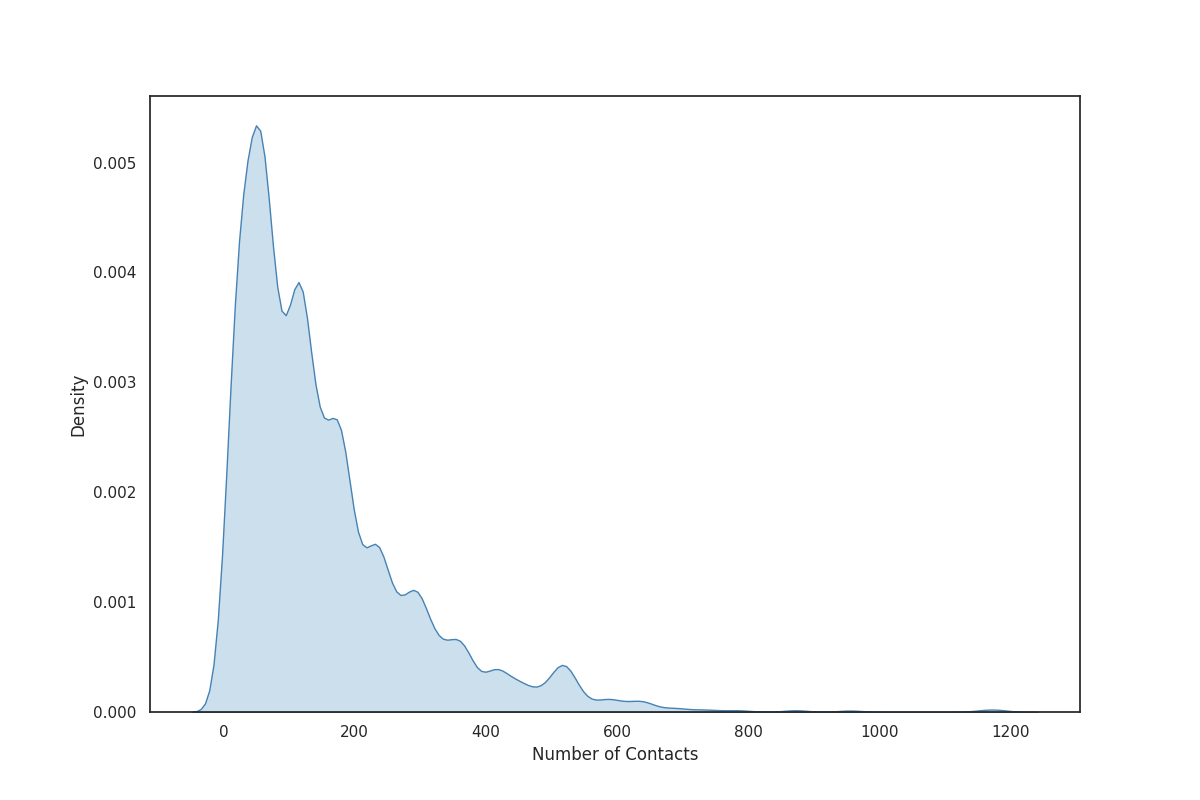}
        \subcaption{Degree distribution}
        \label{deg}
    \end{minipage}
    \hspace{0.01\linewidth}
    \begin{minipage}[b]{0.30\linewidth}
        \centering
        \includegraphics[width=\linewidth]{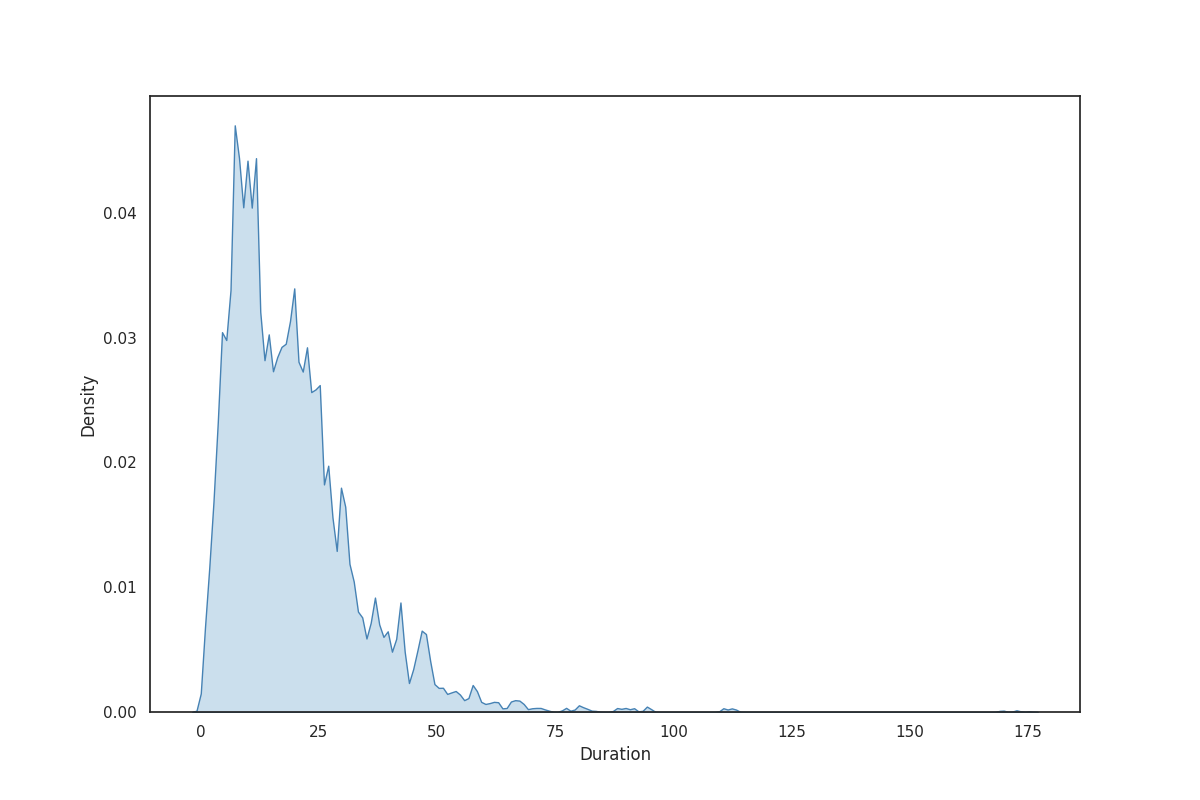}
        \subcaption{Distribution of contact duration (minutes)}
        \label{dur}
    \end{minipage}
    
    \caption{The distribution of (a) contact stat times, (b) clique sizes, (c) the degree distribution and (d) distribution of contact duration}
\end{figure}

For the baseline demand generated by the SF-CHAMP tool, the contact network constructed from the FAST-TrIPs outputs consisted of 48,546 passenger nodes and 3,756,340 contact links. Figure \ref{cntct} shows the network's density plot of contact start times. It can be seen that the traffic peaks around 7 a.m. and 5 p.m., reflecting a typical weekday commute pattern: people use public transportation to go to work in the morning and travel home in the late afternoon. The network forms a fully connected subgraph or clique when passengers travel together between stations on the same vehicle. The distribution of the sizes of these patterns can be seen in Figure \ref{clique}.

The distribution is heavily right-skewed, with most cliques having a relatively small size (below 30). There are two notable peaks, one between clique sizes 0 and 20 and another around 30. Significantly few cliques extend beyond the size 50. This suggests that significant passenger interactions are typically localized, with occasional larger groupings that may occur on high-capacity vehicles (such as trains). The network's degree distribution, which follows a skewed power law with an average degree of 154 and a maximum degree of 1,193, is shown in Figure \ref{deg}. In contrast, Figure \ref{dur} displays the distribution of contact durations. Most contacts last several minutes, with an average contact duration of 19 minutes and 13 seconds.

\subsection*{Design of Transit Scenarios}

To simulate a variety of potential behaviors during an epidemic outbreak, we defined multiple scenarios that changed the original public transit network and demand. The baseline simulation used the unchanged transit demand and vehicle capacities. Our demand contained 117,000 passenger trips done by 52,816 passengers. To define realistic scenarios regarding the demand changes, we used the \href{https://www.google.com/covid19/mobility/}{Google Mobility Report} for September 2022 from California. Based on the report, the average decrease in mobility trends regarding the public transportation hubs (such as subways, buses, train stations, etc.) was $-33\%$ compared to the baseline, which means during a COVID peak, $33\%$ fewer passengers used the transit system than before. However, mobility trends fluctuated throughout the month. Therefore, we created four new demand scenarios excluding $17\%$, $33.5\%$, $41\%$ and $50\%$ of the passengers from the original demand, along with all their associated trips. To model social distancing measures, we also decreased the capacity of each vehicle in our public transit system, reducing the capacities in the GTFS PLUS descriptions by $10\%$, $20\%$, $30\%$, and $50\%$, resulting in four different transit supply inputs. Each pair of demand and supply files defines a single use case, resulting in 20 different scenarios altogether (21 with the baseline scenario), highlighting different outcomes with decreased demand and possible epidemiological measures. The following table introduces the different scenarios in detail.

% 
%Corrections needed due to new baseline demand--------------
%\begin{itemize}
%    \item new baseline demand: 117,000 trips 52,816 passengers
%    \item According to Google Mobility Report average decresase for SF Bay %Area is around 33\% (calculated as average of averages provided for each %county in Bay Area)
%    \item We now have 4 demand files with -17\%,-33.5\%,-41\%,-50\% trips %compared to the baseline; Can we say, since we created new demand %scenarios by excluding passengers, that new scenarios correspond to %7\%,33\%,41\%,50\% decrease to original demand?
%   \item We now have 21 scenario altogether (4x5 + original)
%end{itemize}

The study\cite{park2021risk} describes a model for estimating the risk of COVID-19 infection in public transportation, considering factors such as exposure time, mask efficiency, ventilation rate, and distance. It demonstrates that infection risk lowers as the number of passengers in the vehicle decreases, with the risk heavily influenced by passenger arrangement within the vehicle. Another study\cite{chen2021} estimates the COVID-19 transmission probability using infection data from the Diamond Princess Cruise Ship, Monte Carlo simulations, and an improved Wells-Riley model. This study calculates the reproductive number $R$, which represents the average number of secondary infections caused by a primary infected individual during their infectious period. The value of $R$ for a 48-seat bus during a 2-hour ride was estimated to be 7.48. It was also found that $R$ decreases almost linearly as the number of passengers decreases; for instance, when capacity is reduced to 24 passengers, $R$ is estimated to be 6.57.

In the epidemiological scenarios we designed, we also aimed to incorporate changes in transmission probability due to vehicle capacity reductions. Due to the lack of detailed information about parameters related to specific transit vehicles, such as ventilation rates and seating arrangements, the variation in passenger numbers across different vehicles and times, and the focus of our study on virus transmission in buses, we opted to use simple heuristics for our approach. We divided the reproductive numbers estimated in \cite{chen2021} for a 2-hour ride in a 48-seat bus, under various capacity constraints, by the total number of seats in an average bus to assess the transmission probability. Consequently, we set the value of $P_{max}$ to 0.163 for the scenario with unchanged vehicle capacity and to 0.16, 0.158, 0.156, and 0.14 for scenarios with capacity reductions of 10\%, 20\%, 30\%, and 50\%, respectively. The exact values are in Table \ref{tab:probabilies}.

\begin{table}[ht]
\centering
\begin{tabular}{|c|c|}
\hline
Capacity         & Infection Probability ranges on the Edges \\
\hline
100\% capacity    & 0\% - 16.3\% \\
\hline
90\% capacity    & 0\% - 16.0\% \\
\hline
80\% capacity    & 0\% - 15.8\% \\
\hline
70\% capacity    & 0\% - 15.6\% \\
\hline
50\% capacity    & 0\% - 14\% \\
\hline
\end{tabular}
\caption{\label{tab:probabilies}Infection probability under different capacity reductions.}
\end{table}

\section*{Results and discussion}

In presenting our results, we focus on the effects of reduced demand and vehicle capacity scenarios on passenger interactions. First, we highlight the main changes in the passenger's behavior, describing the differences in passenger travel patterns. In the second part, we focus on the epidemiological effects, trying to find which scenario had the best effect on the number of infected passengers. At the same time, at the end of the result section, we identify the critical parts of the system from an epidemiological point of view.

\subsection*{Effects of reduced demand and vehicle capacity}

As discussed in previous sections, the contact network describes the fine structure of passenger co-traveling relations during the day. Table \ref{tab:graph_stats} summarizes the changes in the key network metrics of the contact network in the case of different demand and vehicle capacity-related restrictions. These metrics include the number of passengers (nodes), the number of connections (edges), as well as max, median, and mean degree (co-traveling connections of individual passengers). If a vehicle travels between stations A and B, the passengers sharing the trip will form a clique (fully connected subgraph) in the network. Therefore, metrics also include max, mean, and median clique sizes describing the sizes of co-traveling groups between stations. The results in the table suggest that capacity restrictions and demand-related changes can effectively disrupt the formation of large high-risk clusters, potentially reducing the risk related to infectious passengers within the transit network.

The results show a significant decrease in the overall size of the passenger contact network (as shown by the number of nodes and edges) through the different scenarios. These changes highlight the impact of the demand and vehicle capacity changes on the network structure. 

\begin{table}[ht]
\centering
\begin{tabular}{|c|c|c|c|c|c|c|c|c|c|}
\hline
Demand & Vehicle & Max & Median & Mean & Max Clique & Median Clique & Mean Clique & Number of & Number of \\
& Capacity & Degree & Degree & Degree & Size & Size & Size & Nodes & Edges \\
\hline
100.0\% & 100.0\% & 1193 & 118 & 154.754 & 333 & 26 & 23.640 & 48546 & 3756340 \\
\hline
100.0\% & 90.0\% & 1151 & 112 & 146.649 & 330 & 27 & 24.184 & 47753 & 3501454 \\
\hline
100.0\% & 80.0\% & 1063 & 100 & 139.102 & 328 & 26 & 23.904 & 46837 & 3257555 \\
\hline
100.0\% & 70.0\% & 973 & 88 & 131.529 & 328 & 25 & 23.641 & 45833 & 3014174 \\
\hline
100.0\% & 50.0\% & 968 & 77 & 119.289 & 328 & 26 & 25.723 & 43613 & 2601283 \\
\hline
83.0\% & 90.0\% & 1009 & 100 & 132.150 & 278 & 18 & 20.515 & 41571 & 2746795 \\
\hline
83.0\% & 80.0\% & 1005 & 99 & 128.081 & 278 & 18 & 20.334 & 40973 & 2623940 \\
\hline
83.0\% & 70.0\% & 928 & 88 & 119.889 & 273 & 18 & 20.081 & 40130 & 2405574 \\
\hline
83.0\% & 50.0\% & 839 & 69 & 105.415 & 273 & 21 & 20.818 & 38035 & 2004730 \\
\hline
66.5\% & 90.0\% & 799 & 80 & 109.814 & 212 & 12 & 16.668 & 33330 & 1830045 \\
\hline
66.5\% & 80.0\% & 799 & 80 & 107.656 & 212 & 13 & 16.595 & 33147 & 1784235 \\
\hline
66.5\% & 70.0\% & 797 & 78 & 105.273 & 212 & 13 & 16.374 & 32894 & 1731428 \\
\hline
66.5\% & 50.0\% & 642 & 63 & 91.301 & 212 & 13 & 15.632 & 31177 & 1423242 \\
\hline
59.0\% & 90.0\% & 681 & 72 & 97.447 & 188 & 17 & 17.968 & 29313 & 1428233 \\
\hline
59.0\% & 80.0\% & 681 & 70 & 95.523 & 188 & 15 & 17.292 & 29134 & 1391489 \\
\hline
59.0\% & 70.0\% & 681 & 70 & 94.011 & 188 & 16 & 17.450 & 28931 & 1359921 \\
\hline
59.0\% & 50.0\% & 637 & 62 & 84.768 & 188 & 16 & 17.416 & 27893 & 1182214 \\
\hline
50.0\% & 90.0\% & 548 & 64 & 85.525 & 173 & 13 & 16.245 & 25198 & 1077533 \\
\hline
50.0\% & 80.0\% & 548 & 62 & 84.273 & 173 & 13 & 16.192 & 25107 & 1057925 \\
\hline
50.0\% & 70.0\% & 548 & 62 & 82.304 & 173 & 13 & 16.134 & 24974 & 1027731 \\
\hline
50.0\% & 50.0\% & 548 & 58 & 77.067 & 171 & 13 & 15.420 & 24376 & 939292 \\
\hline
\end{tabular}
\caption{\label{tab:graph_stats}Contact network statistics through the different demand and vehicle capacity scenarios.}
\end{table}

The restrictions also impacted the degree distribution of the contact network. As the degree of a given node represents the number of direct contacts a passenger has during the day, the risk of exposure to infectious diseases is reduced as the demand and capacities decrease. At 100\% demand, the degree of the passenger with the highest number of connections reached as high as 1193; through the different scenarios, it dropped to 548, together with the median and mean number of co-travelings. 

Diseases spread more quickly inside a community or a strongly connected cluster due to the higher number of connections among the passengers. Therefore, as passenger groups grow and become more interconnected in the network, they promote the spreading process, working as supers spreaders during the epidemiological simulation. Consecutively, clique sizes and clique size distributions are expected to correlate highly with the final epidemiological modeling results.

\subsection*{Demand serving efficiency of the transit system}

The structure of the transit system does not always match the given structure of the demand due to the lack of infrastructure. A natural example is if too many passengers are trying to reach a destination simultaneously, and only a limited number of vehicle trips are going toward the station. If the system can not handle individual passengers due to the infeasibility of their demand, they cannot reach their destination, and the transit demand model leaves them out of the simulation and the contact network simultaneously. Therefore, for each scenario, we have a small portion of passengers that could not satisfy their demand due to the vehicle capacity restrictions. In this way, critical parts of the system can be identified from an efficiency point of view, designating vehicle trips that are sensitive to the reduced capacities, leaving passengers without transit options. This information can also be used to evaluate the system's capability to serve passengers. Obviously, as seen in Table \ref{tab:notreaching}, the number of passengers failing to reach their destinations increases as vehicle capacity restrictions are introduced. Consequently, reducing demand naturally reduces the load on sensitive vehicle systems, helping the system to be efficient.

\begin{table}[ht]
\centering
\begin{tabular}{|c|c|c|c|c|}
\hline
Original 4,270 & 90\% capacity& 80\% capacity& 70\% capacity& 50\% capacity\\
\hline
100\% demand  & 5,063 & 5,979 & 6,983 & 9,203 \\
\hline
83\% demand   & 3,341 & 3,939 & 4,782 & 6,877 \\
\hline
66.5\% demand & 2,150 & 2,333 & 2,586 & 4,303 \\
\hline
59\% demand   & 1,676 & 1,855 & 2,058 & 3,096 \\
\hline
50\% demand   & 1,300 & 1,391 & 1,524 & 2,122 \\
\hline
\end{tabular}
\caption{\label{tab:notreaching}Number of passengers failing to reach their destination during the transit demand model simulation}
\end{table}

Under normal operating conditions, with 100\% demand and unrestricted vehicle capacity, 4270 passengers fail to reach their destination. We use this as a baseline to compare the effects of the different scenarios. Table \ref{tab:notreaching} shows that as capacity restrictions are introduced, the number of passengers with infeasible demand grows substantially. The trend gets worse as the capacity restrictions become more strict, reaching as far as 9203 stranded passengers at the end of the restriction range. This means that introducing only transit system-related restrictions significantly damages efficiency. However, on the other hand, even a slight decrease in demand helps ease the load on critical components, helping the system to serve more and more passenger demand.

The results demonstrate the necessary balance between capacity restrictions and demand management. Combining different demands and vehicle capacity-reducing provisions can help maintain operational efficiency while protecting passengers in the case of an outbreak.

\subsection*{Epidemiological Spread Analysis}

In this section, we overview the results of the epidemiological simulation focusing on global rate of infected passengers and the number of highly endangered individuals. We also provide these values as a ratio compared to the network size in each scenario to compare the real epidemiological effects of the restrictions. 

At the beginning of each simulation process, we started with 100 infected passengers ($|A_0| =100$), selected randomly from the network, and then we simulated 5 iterations representing the infection period of 5 days. We ran the epidemic spreading process $k = 100000$ times to estimate the infection probabilities. As we discussed before, we selected $\tau_i$ so that no nodes transitioned to the removed state, remaining infectious throughout the simulation.

The epidemiological simulation shows that reduced demand and vehicle capacities affect the global infection rate of the contact network. Table \ref{tab:globalinfection} presents the expected percentage of passengers getting infected at the end of the simulation.

\begin{table}[ht]
\centering
\begin{tabular}{|c|c|c|c|c|}
\hline
Original: 30.94\% & 90\% capacity & 80\% capacity & 70\% capacity & 50\% capacity \\
\hline
100\% demand  & 30.34\% & 30.04\% & 29.66\% & 27.90\% \\
\hline
83\% demand & 28.36\% & 28.02\% & 27.59\% & 25.68\% \\
\hline
66.5\% demand & 25.54\% & 25.16\% & 24.80\% & 22.68\% \\
\hline
59\% demand & 23.78\% & 23.44\% & 23.28\% & 21.06\% \\
\hline
50\% demand & 22.23\% & 21.93\% & 21.59\% & 19.43\% \\
\hline
\end{tabular}
\caption{\label{tab:globalinfection}Percentage of infected passengers at the end of the simulation corresponding to the 21 use cases.}
\end{table}

The baseline scenario, where we did not have reduced demand or restricted capacities, shows an infection rate of 30.94\%. As the capacity restriction becomes more strict, global infection slightly declines, reducing to 27.9\%. Similar but more pronounced trend can be observed for lower demand levels. At different demand steps, infection declines more and more, reaching 22.23\% at 50\% demand reduction. Results show that the reduction of infection rate is more noticeable in the case of demand change than in restricted capacities. However, implementing capacity restrictions, even without significantly reducing passenger demand, can lower infection rates within the transit network. Furthermore, the combined effect of restrictions regarding the transit system and governmental regulations that affect the demand can substantially reduce the impact of an outbreak on society. The results underscore the importance of coordinated measures in managing public health risks during pandemics.

 To find out the changes in the number of infected passengers in a more refined way, we collected highly endangered passengers -- those with a final infection probability greater than 50\% -- from the different demand and capacity-reduced scenarios. The table \ref{tab:dangered} provides a different view of the results.

\begin{table}[ht]
\centering
\begin{tabular}{|c|c|c|c|c|}
\hline
Original 12,789 & 90\% capacity & 80\% capacity & 70\% capacity & 50\% capacity \\
\hline
100\% demand  & 12331 & 12113 & 11767 & 10,540 \\
\hline
83\% demand   & 9859 & 9660 & 9358 & 8220 \\
\hline
66.5\% demand & 6920 & 6758 & 6590 & 5659 \\
\hline
59\% demand   & 5446 & 5349 & 5232 & 4597 \\
\hline
50\% demand   & 4307 & 4226 & 4132 & 3630 \\
\hline
\end{tabular}
\caption{\label{tab:dangered}Number of highly endangered passengers along the different scenarios. (Where the final infection probability of the passenger was greater than 50\%). }
\end{table}

In the original scenario, with full capacity and demand, capacity restrictions reduce the number of highly endangered individuals from 12789 to 10540 by 17.5\%. The results show the same trend as the global infection percentage, so as the demand decreases and capacity restrictions are introduced, the number of passengers declines even further. Changing capacity restrictions from 0\% to 50 \%, the number drops to 4307, lowering the number of endangered passengers by 66.3\%. In another example, when decision-makers make 50\% capacity restrictions together with local measures that reduce the demand by 50\%, the number of critical infection level individuals drops by 71\%. 

The results highlight that demand-related restrictions have a higher effect on the infection rate. Although capacity restrictions are essential parts of containment strategies, they are not necessarily enough to minimize the negative impact of an outbreak situation. Finally, combining transit system-related measures with demand-reducing decisions can maximize the reduction in infected individuals.

%MARK

\subsection*{Identifying critical components}

Bota et al.\cite{bota2017b_modeling} proposed a methodology for identifying the most critical bus trips responsible for disease spreading among passengers. Here we follow the same approach by taking the average infection probabilities for all passengers on a particular vehicle trip to estimate the likelihood of transporting infected passengers. From another perspective, this gives us an estimate of the probability of getting infected if a passenger uses that specific vehicle trip. To calculate the most critical parts of the system, we take the top 100 vehicle trips based on this metric, according to the highest infection likelihoods.

Since we are interested in measuring the changes caused by our scenarios in the epidemiological danger of public transportation system components, we took a slightly different approach. Instead of focusing on individual vehicle trips, we evaluated a similar likelihood of infection but associated with transit routes as a whole (i.e., any vehicle traveling on a particular route within a 24-hour time frame). In the following part of the paper, we summarize the results for the top 10 most critical bus routes identified using this approach.

\begin{figure}[ht]
\centering
\includegraphics[width=0.4\linewidth]{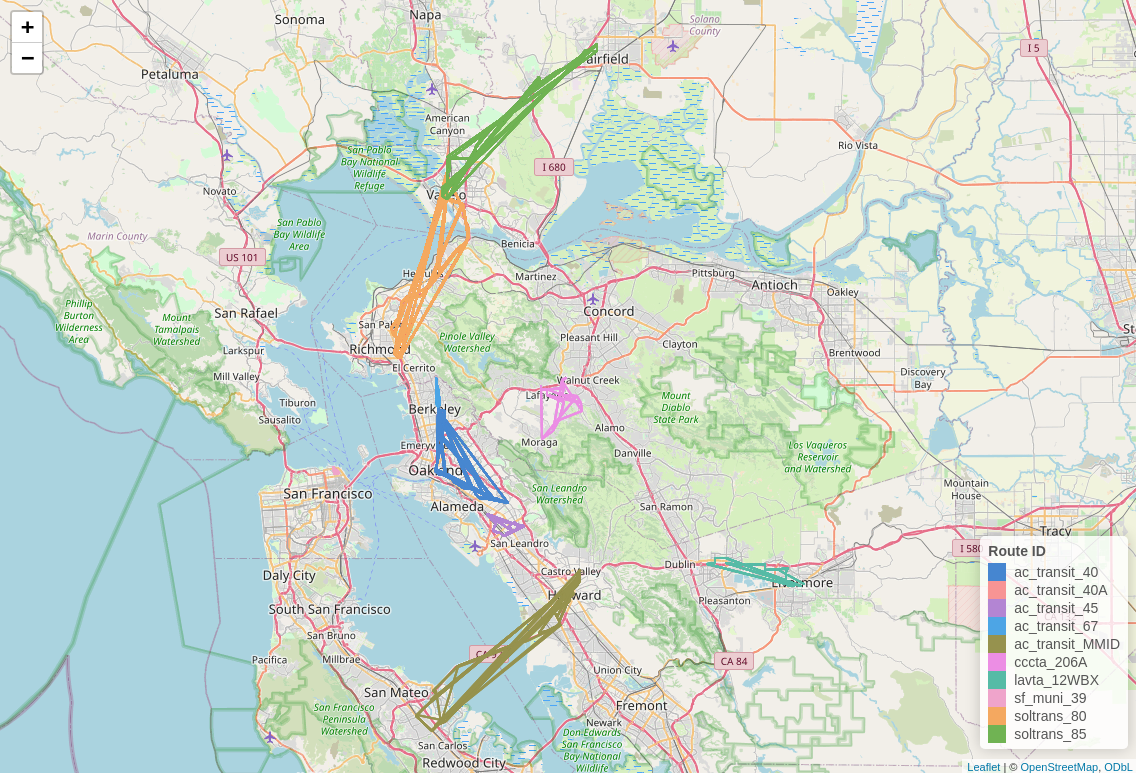}
\caption{Critical regions in the transit system}
\label{fig:pubnetwork2}
\end{figure}

Figure \ref{fig:pubnetwork2} shows the most critical bus routes and the regions they connect. As it can be seen, these transit routes are located in different parts of the Bay Area region, and most of them connect different counties or cities. However, there are other routes, for example, cccta\_206A, operated by the Central Contra Costa Transit Authority, or sf\_muni\_39, which operates within San Francisco. Most critical routes include the connection between Fremont and Palo Alto, Richmond and Vallejo, Vallejo and Fairfield, or Berkeley and San Leandro. Here, if passengers use these routes, they get infected with a likelihood higher than 90\% in the case of the baseline scenario. In the following, we evaluate how the different demand and vehicle capacity-related restrictions affect these routes.

\begin{figure}[!ht]
    \centering
    \begin{subfigure}[b]{0.49\linewidth}
        \centering
        \includegraphics[width=\linewidth]{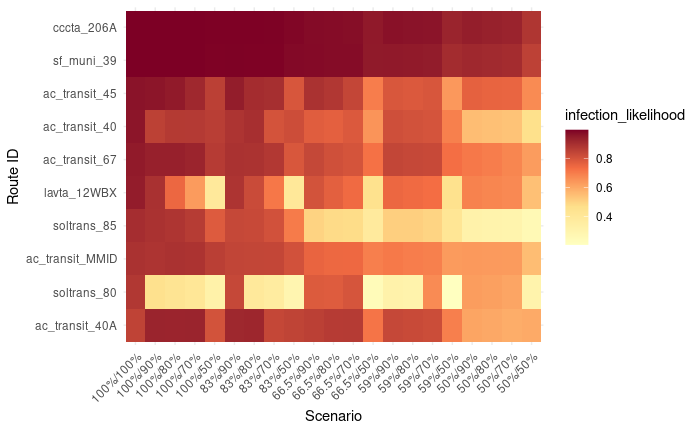}
        \label{fig:criticalhm}
    \end{subfigure}
    \hfill
    \begin{subfigure}[b]{0.49\linewidth}
        \centering
        \includegraphics[width=\linewidth]{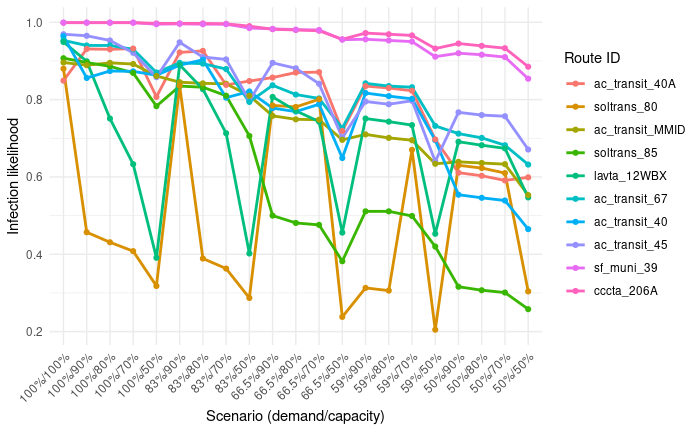}
        \label{fig:second_image}
    \end{subfigure}
    \caption{Effect of the different scenarios along the top 10 most contagious vehicles. The x-axis shows the demand and capacity percentages of the different scenarios. }
    \label{fig:two_critical}
\end{figure}

Figure \ref{fig:two_critical} shows changes in the expected infection likelihood in the scenarios with different vehicle capacities and travel demand reductions. It can be seen that the infection probabilities decrease as transit demand lowers for all identified routes. Among the top 10 routes, two types of behaviors can be observed. The first is when the observed infection likelihoods decrease slightly along the different scenarios, for example, in the case of sf\_muni\_39 or cccta\_206A. However, interesting trends could be observed in the rest of the routes, especially for routes soltrans\_80 that travel between Fairfield and Vallejo cities and lavta\_12WBX belonging to Wheel public bus service operating in the Tri-Valley region. We can spot the drastic impact of vehicle capacities on the likelihood of infection for these routes. The restrictions cause a 50-70\% decrease in expected infection probabilities caused only by vehicle capacity restrictions. It can be seen that in some cases, demand-related changes did not affect the route-related risks. In other cases, these restrictions worked best to reduce the disease-related dangers. Results show that some routes react differently to the various types of restrictions, highlighting the importance of combining multiple measures. This strategy will balance the reduction effect in the different parts of the system.

\section*{Limitations}

Due to the nature of the simulation environment, the main limitation of the study is the size of the demand data. Although the GTFS structure is defined in a way that it serves hundreds of thousands on a daily basis, computational constraints and the transit model prevented us from simulating the real-world demand on the transit network. As a result, long train routes were not filled beyond half capacity, making it impossible to simulate the disease transmission on the trains realistically. Therefore, the results introduced in this research focus on buses where the simulation was closer to a realistic scenario. It is also important to note that we aimed to show the effects of the different measures, focusing on the differences between the baseline and reduced scenarios. Our proposed method is proof of concept but can be directly applied to more comprehensive travel data sets in future works.

\section*{Conclusions}

Based on the results, it can be observed that social distancing, meaning introducing vehicle capacity restrictions, definitely impacts the epidemiological spread, lowering the risk in the system. However, governmental decisions that try to reduce the demand have more significant potential to decrease the percentage of final infected passengers and highly infected individuals. Introducing only transit system-related measures also risks the system's efficiency, resulting in a greater number of passengers who cannot reach their destination using only the public transportation system. Due to these factors and the simulation results, we can conclude that combining demand-reducing governmental measures with transit system-related restrictions is the best solution. This solution minimizes the epidemiological risk while maintaining the system's service efficiency.

\section*{Data availability statement}

The GTFS and demand input files will be available on Zenodo.org upon publication.
The input files as well as the source code is also available on request from the corresponding author.

\section*{Competing interests}
The authors declare no competing interests.

\section*{Acknowledgements}

László Hajdu acknowledges the European Commission for funding the InnoRenew CoE project (Grant Agreement no. 739574) under the Horizon2020 Widespread-Teaming program and the Republic of Slovenia (Investment funding of the Republic of Slovenia and the European Union of the European Regional Development Fund). He is also grateful for the support of the University of Primorska through postdoc grant No. 2991-10/2022.

\section*{Author contributions statement}

Concept and design: LH, AB, MK Data acquisition: JP, LH, Methodology: LH, AB, MK, Data analysis: JP, LH, Software: JP, LH, Writing: LH, AB, JP.

\bibliography{sample}

\begin{thebibliography}{10}
\urlstyle{rm}
\expandafter\ifx\csname url\endcsname\relax
  \def\url#1{\texttt{#1}}\fi
\expandafter\ifx\csname urlprefix\endcsname\relax\def\urlprefix{URL }\fi
\expandafter\ifx\csname doiprefix\endcsname\relax\def\doiprefix{DOI: }\fi
\providecommand{\bibinfo}[2]{#2}
\providecommand{\eprint}[2][]{\url{#2}}

\bibitem{tirachini2020covid}
\bibinfo{author}{Tirachini, A.} \& \bibinfo{author}{Cats, O.}
\newblock \bibinfo{journal}{\bibinfo{title}{Covid-19 and public transportation: Current assessment, prospects, and research needs}}.
\newblock {\emph{\JournalTitle{Journal of Public Transportation}}} \textbf{\bibinfo{volume}{22}}, \bibinfo{pages}{1--21}, \doiprefix\url{10.5038/2375-0901.22.1.1} (\bibinfo{year}{2020}).
\newblock \bibinfo{note}{Epub 2022 Sep 13}.

\bibitem{GKIOTSALITIS2021374}
\bibinfo{author}{Gkiotsalitis, K.} \& \bibinfo{author}{Cats, O.}
\newblock \bibinfo{journal}{\bibinfo{title}{Public transport planning adaption under the covid-19 pandemic crisis: literature review of research needs and directions}}.
\newblock {\emph{\JournalTitle{Transport Reviews}}} \textbf{\bibinfo{volume}{41}}, \bibinfo{pages}{374--392}, \doiprefix\url{https://doi.org/10.1080/01441647.2020.1857886} (\bibinfo{year}{2021}).

\bibitem{Kerr}
\bibinfo{author}{Kerr, C.} \emph{et~al.}
\newblock \bibinfo{journal}{\bibinfo{title}{Covasim: An agent-based model of covid-19 dynamics and interventions}}.
\newblock {\emph{\JournalTitle{PLOS Computational Biology}}} \textbf{\bibinfo{volume}{17}}, \bibinfo{pages}{e1009149}, \doiprefix\url{10.1371/journal.pcbi.1009149} (\bibinfo{year}{2021}).

\bibitem{apple_mobility_2020}
\bibinfo{author}{{Apple Inc.}}
\newblock \bibinfo{title}{Apple mobility trends reports} (\bibinfo{year}{2020}).
\newblock \bibinfo{note}{Retrieved from https://www.apple.com/covid19/mobility}.

\bibitem{google_mobility_2020}
\bibinfo{author}{{Google LLC}}.
\newblock \bibinfo{title}{Google covid-19 community mobility reports} (\bibinfo{year}{2020}).
\newblock \bibinfo{note}{Retrieved from https://www.google.com/covid19/mobility/}.

\bibitem{Brauer2008}
\bibinfo{author}{Brauer, F.}
\newblock \emph{\bibinfo{title}{Compartmental Models in Epidemiology}}, \bibinfo{pages}{19--79} (\bibinfo{publisher}{Springer Berlin Heidelberg}, \bibinfo{address}{Berlin, Heidelberg}, \bibinfo{year}{2008}).

\bibitem{kempe}
\bibinfo{author}{Kempe, D.}, \bibinfo{author}{Kleinberg, J.} \& \bibinfo{author}{Tardos, E.}
\newblock \bibinfo{journal}{\bibinfo{title}{Maximizing the spread of influence through a social network}}.
\newblock {\emph{\JournalTitle{Proceedings of the ACM SIGKDD International Conference on Knowledge Discovery and Data Mining}}} \textbf{\bibinfo{volume}{137-146}}, \doiprefix\url{10.1145/956750.956769} (\bibinfo{year}{2003}).

\bibitem{granovetter}
\bibinfo{author}{Granovetter, M.}
\newblock \bibinfo{journal}{\bibinfo{title}{Threshold models of collective behavior}}.
\newblock {\emph{\JournalTitle{The American Journal of Sociology}}} \textbf{\bibinfo{volume}{83}}, \bibinfo{pages}{1420--1443} (\bibinfo{year}{1978}).

\bibitem{walter2012compartmental}
\bibinfo{author}{Walter, G.} \& \bibinfo{author}{Contreras, M.}
\newblock \emph{\bibinfo{title}{Compartmental Modeling with Networks}}.
\newblock Modeling and Simulation in Science, Engineering and Technology (\bibinfo{publisher}{Birkh{\"a}user Boston}, \bibinfo{year}{2012}).

\bibitem{Bóta_Krész_Pluhár_2013}
\bibinfo{author}{Bóta, A.}, \bibinfo{author}{Krész, M.} \& \bibinfo{author}{Pluhár, A.}
\newblock \bibinfo{journal}{\bibinfo{title}{Approximations of the generalized cascade model}}.
\newblock {\emph{\JournalTitle{Acta Cybernetica}}} \textbf{\bibinfo{volume}{21}}, \bibinfo{pages}{37--51}, \doiprefix\url{10.14232/actacyb.21.1.2013.4} (\bibinfo{year}{2013}).

\bibitem{coleman1996medical}
\bibinfo{author}{Coleman, J.}, \bibinfo{author}{Menzel, H.} \& \bibinfo{author}{Katz, E.}
\newblock \emph{\bibinfo{title}{Medical Innovations: A Diffusion Study}} (\bibinfo{publisher}{Bobbs Merrill}, \bibinfo{address}{New York, NY}, \bibinfo{year}{1996}).

\bibitem{hasan2011contagion}
\bibinfo{author}{Hasan, S.} \& \bibinfo{author}{Ukkusuri, S.~V.}
\newblock \bibinfo{journal}{\bibinfo{title}{A contagion model for understanding the propagation of hurricane warning information}}.
\newblock {\emph{\JournalTitle{Transportation Research Part B: Methodological}}} \textbf{\bibinfo{volume}{45}}, \bibinfo{pages}{1590--1605} (\bibinfo{year}{2011}).

\bibitem{sun2014efficient}
\bibinfo{author}{Sun, L.}, \bibinfo{author}{Axhausen, K.~W.}, \bibinfo{author}{Lee, D.~H.} \& \bibinfo{author}{Cebrian, M.}
\newblock \bibinfo{journal}{\bibinfo{title}{Efficient detection of contagious outbreaks in massive metropolitan encounter networks}}.
\newblock {\emph{\JournalTitle{Scientific Reports}}} \textbf{\bibinfo{volume}{4}}, \bibinfo{pages}{5099}, \doiprefix\url{10.1038/srep05099} (\bibinfo{year}{2014}).

\bibitem{Lam2003}
\bibinfo{author}{Lam, W.} \& \bibinfo{author}{Huang, H.}
\newblock \bibinfo{journal}{\bibinfo{title}{Combined activity/travel choice models: Time-dependent and dynamic versions}}.
\newblock {\emph{\JournalTitle{Networks and Spatial Economics}}} \textbf{\bibinfo{volume}{3}}, \bibinfo{pages}{323--347} (\bibinfo{year}{2003}).

\bibitem{Roorda2009}
\bibinfo{author}{Roorda, M.}, \bibinfo{author}{Carrasco, J.} \& \bibinfo{author}{Miller, E.}
\newblock \bibinfo{journal}{\bibinfo{title}{An integrated model of vehicle transactions, activity scheduling and mode choice}}.
\newblock {\emph{\JournalTitle{Transportation Research Part B}}} \textbf{\bibinfo{volume}{43}}, \bibinfo{pages}{217--229} (\bibinfo{year}{2009}).

\bibitem{Sun2013}
\bibinfo{author}{Sun, L.}, \bibinfo{author}{Axhausen, K.}, \bibinfo{author}{Lee, D.} \& \bibinfo{author}{Huang, X.}
\newblock \bibinfo{journal}{\bibinfo{title}{Understanding metropolitan patterns of daily encounters}}.
\newblock {\emph{\JournalTitle{Proceedings of the National Academy of Sciences of the United States of America}}} \textbf{\bibinfo{volume}{110}}, \bibinfo{pages}{13774--13779}, \doiprefix\url{10.1073/pnas.1306440110} (\bibinfo{year}{2013}).
\newblock \bibinfo{note}{PMCID: PMC3752247, PMID: 23918373}.

\bibitem{Bta2011SystematicLO}
\bibinfo{author}{B{\'o}ta, A.}, \bibinfo{author}{Kr{\'e}sz, M.} \& \bibinfo{author}{Pluh{\'a}r, A.}
\newblock \bibinfo{title}{Systematic learning of edge probabilities in the domingos-richardson model} (\bibinfo{year}{2011}).

\bibitem{6932999}
\bibinfo{author}{Bóta, A.}, \bibinfo{author}{Krész, M.} \& \bibinfo{author}{Pluhár, A.}
\newblock \bibinfo{title}{The inverse infection problem}.
\newblock In \emph{\bibinfo{booktitle}{2014 Federated Conference on Computer Science and Information Systems}}, \bibinfo{pages}{75--84}, \doiprefix\url{10.15439/2014F261} (\bibinfo{year}{2014}).

\bibitem{cattuto2010dynamics}
\bibinfo{author}{Cattuto, C.} \emph{et~al.}
\newblock \bibinfo{journal}{\bibinfo{title}{Dynamics of person-to-person interactions from distributed rfid sensor networks}}.
\newblock {\emph{\JournalTitle{PLOS ONE}}} \textbf{\bibinfo{volume}{5}}, \bibinfo{pages}{e11596}, \doiprefix\url{10.1371/journal.pone.0011596} (\bibinfo{year}{2010}).

\bibitem{christakis2010social}
\bibinfo{author}{Christakis, N.~A.} \& \bibinfo{author}{Fowler, J.~H.}
\newblock \bibinfo{journal}{\bibinfo{title}{Social network sensors for early detection of contagious outbreaks}}.
\newblock {\emph{\JournalTitle{PLOS ONE}}} \textbf{\bibinfo{volume}{5}}, \bibinfo{pages}{e12948}, \doiprefix\url{10.1371/journal.pone.0012948} (\bibinfo{year}{2010}).

\bibitem{2020_Buchanan}
\bibinfo{author}{Buchanan, W.~J.} \emph{et~al.}
\newblock \bibinfo{journal}{\bibinfo{title}{Review and critical analysis of privacy-preserving infection tracking and contact tracing}}.
\newblock {\emph{\JournalTitle{Frontiers in Communications and Networks}}} \textbf{\bibinfo{volume}{1}}, \doiprefix\url{10.3389/frcmn.2020.583376} (\bibinfo{year}{2020}).

\bibitem{collecting}
\bibinfo{author}{et. al., T.}
\newblock \bibinfo{journal}{\bibinfo{title}{Human mobility data in the covid-19 pandemic: characteristics, applications, and challenges}}.
\newblock {\emph{\JournalTitle{International Journal of Digital Earth}}} \textbf{\bibinfo{volume}{14}}, \bibinfo{pages}{1126--1147}, \doiprefix\url{10.1080/17538947.2021.1952324} (\bibinfo{year}{2021}).

\bibitem{MO2021102893}
\bibinfo{author}{Mo, B.} \emph{et~al.}
\newblock \bibinfo{journal}{\bibinfo{title}{Modeling epidemic spreading through public transit using time-varying encounter network}}.
\newblock {\emph{\JournalTitle{Transportation Research Part C: Emerging Technologies}}} \textbf{\bibinfo{volume}{122}}, \bibinfo{pages}{102893}, \doiprefix\url{https://doi.org/10.1016/j.trc.2020.102893} (\bibinfo{year}{2021}).

\bibitem{Arenas2020.03.21.20040022}
\bibinfo{author}{Arenas, A.} \emph{et~al.}
\newblock \bibinfo{journal}{\bibinfo{title}{A mathematical model for the spatiotemporal epidemic spreading of covid19}}.
\newblock {\emph{\JournalTitle{medRxiv}}} \doiprefix\url{10.1101/2020.03.21.20040022} (\bibinfo{year}{2020}).
\newblock \eprint{https://www.medrxiv.org/content/early/2020/03/23/2020.03.21.20040022.full.pdf}.

\bibitem{PARE2020345}
\bibinfo{author}{Paré, P.~E.}, \bibinfo{author}{Beck, C.~L.} \& \bibinfo{author}{Başar, T.}
\newblock \bibinfo{journal}{\bibinfo{title}{Modeling, estimation, and analysis of epidemics over networks: An overview}}.
\newblock {\emph{\JournalTitle{Annual Reviews in Control}}} \textbf{\bibinfo{volume}{50}}, \bibinfo{pages}{345--360}, \doiprefix\url{https://doi.org/10.1016/j.arcontrol.2020.09.003} (\bibinfo{year}{2020}).

\bibitem{10.3389/fpubh.2024.1367324}
\bibinfo{author}{Ly, Y.-T.}, \bibinfo{author}{Leuko, S.} \& \bibinfo{author}{Moeller, R.}
\newblock \bibinfo{journal}{\bibinfo{title}{An overview of the bacterial microbiome of public transportation systems—risks, detection, and countermeasures}}.
\newblock {\emph{\JournalTitle{Frontiers in Public Health}}} \textbf{\bibinfo{volume}{12}}, \doiprefix\url{10.3389/fpubh.2024.1367324} (\bibinfo{year}{2024}).

\bibitem{LIU2022100030}
\bibinfo{author}{Liu, X.}, \bibinfo{author}{Kortoçi, P.}, \bibinfo{author}{Motlagh, N.~H.}, \bibinfo{author}{Nurmi, P.} \& \bibinfo{author}{Tarkoma, S.}
\newblock \bibinfo{journal}{\bibinfo{title}{A survey of covid-19 in public transportation: Transmission risk, mitigation and prevention}}.
\newblock {\emph{\JournalTitle{Multimodal Transportation}}} \textbf{\bibinfo{volume}{1}}, \bibinfo{pages}{100030}, \doiprefix\url{https://doi.org/10.1016/j.multra.2022.100030} (\bibinfo{year}{2022}).

\bibitem{PUJANTEOTALORA2023104422}
\bibinfo{author}{Pujante-Otalora, L.}, \bibinfo{author}{Canovas-Segura, B.}, \bibinfo{author}{Campos, M.} \& \bibinfo{author}{Juarez, J.~M.}
\newblock \bibinfo{journal}{\bibinfo{title}{The use of networks in spatial and temporal computational models for outbreak spread in epidemiology: A systematic review}}.
\newblock {\emph{\JournalTitle{Journal of Biomedical Informatics}}} \textbf{\bibinfo{volume}{143}}, \bibinfo{pages}{104422}, \doiprefix\url{https://doi.org/10.1016/j.jbi.2023.104422} (\bibinfo{year}{2023}).

\bibitem{histo1}
\bibinfo{author}{Vinceti, M.} \emph{et~al.}
\newblock \bibinfo{journal}{\bibinfo{title}{{Substantial impact of mobility restrictions on reducing COVID-19 incidence in Italy in 2020}}}.
\newblock {\emph{\JournalTitle{Journal of Travel Medicine}}} \textbf{\bibinfo{volume}{29}}, \bibinfo{pages}{taac081}, \doiprefix\url{10.1093/jtm/taac081} (\bibinfo{year}{2022}).
\newblock \eprint{https://academic.oup.com/jtm/article-pdf/29/6/taac081/45890884/taac081.pdf}.

\bibitem{histo2}
\bibinfo{author}{Oka, T.}, \bibinfo{author}{Wei, W.} \& \bibinfo{author}{Zhu, D.}
\newblock \bibinfo{journal}{\bibinfo{title}{The effect of human mobility restrictions on the covid-19 transmission network in china}}.
\newblock {\emph{\JournalTitle{PLOS ONE}}} \textbf{\bibinfo{volume}{16}}, \bibinfo{pages}{1--16}, \doiprefix\url{10.1371/journal.pone.0254403} (\bibinfo{year}{2021}).

\bibitem{Murano2021}
\bibinfo{author}{Murano, Y.} \emph{et~al.}
\newblock \bibinfo{journal}{\bibinfo{title}{Impact of domestic travel restrictions on transmission of covid-19 infection using public transportation network approach}}.
\newblock {\emph{\JournalTitle{Scientific Reports}}} \textbf{\bibinfo{volume}{11}}, \bibinfo{pages}{3109}, \doiprefix\url{10.1038/s41598-021-81806-3} (\bibinfo{year}{2021}).

\bibitem{FAZIO2022101373}
\bibinfo{author}{Fazio, M.} \emph{et~al.}
\newblock \bibinfo{journal}{\bibinfo{title}{Exploring the impact of mobility restrictions on the covid-19 spreading through an agent-based approach}}.
\newblock {\emph{\JournalTitle{Journal of Transport \& Health}}} \textbf{\bibinfo{volume}{25}}, \bibinfo{pages}{101373}, \doiprefix\url{https://doi.org/10.1016/j.jth.2022.101373} (\bibinfo{year}{2022}).

\bibitem{LUO2022103592}
\bibinfo{author}{Luo, Q.}, \bibinfo{author}{Gee, M.}, \bibinfo{author}{Piccoli, B.}, \bibinfo{author}{Work, D.} \& \bibinfo{author}{Samaranayake, S.}
\newblock \bibinfo{journal}{\bibinfo{title}{Managing public transit during a pandemic: The trade-off between safety and mobility}}.
\newblock {\emph{\JournalTitle{Transportation Research Part C: Emerging Technologies}}} \textbf{\bibinfo{volume}{138}}, \bibinfo{pages}{103592}, \doiprefix\url{https://doi.org/10.1016/j.trc.2022.103592} (\bibinfo{year}{2022}).

\bibitem{10.1371/journal.pone.0260919}
\bibinfo{author}{Tully, M.~A.} \emph{et~al.}
\newblock \bibinfo{journal}{\bibinfo{title}{The effect of different covid-19 public health restrictions on mobility: A systematic review}}.
\newblock {\emph{\JournalTitle{PLOS ONE}}} \textbf{\bibinfo{volume}{16}}, \bibinfo{pages}{1--18}, \doiprefix\url{10.1371/journal.pone.0260919} (\bibinfo{year}{2021}).

\bibitem{KAMGA202125}
\bibinfo{author}{Kamga, C.} \& \bibinfo{author}{Eickemeyer, P.}
\newblock \bibinfo{journal}{\bibinfo{title}{Slowing the spread of covid-19: Review of “social distancing” interventions deployed by public transit in the united states and canada}}.
\newblock {\emph{\JournalTitle{Transport Policy}}} \textbf{\bibinfo{volume}{106}}, \bibinfo{pages}{25--36}, \doiprefix\url{https://doi.org/10.1016/j.tranpol.2021.03.014} (\bibinfo{year}{2021}).

\bibitem{Ayouni2021}
\bibinfo{author}{Ayouni, I.} \emph{et~al.}
\newblock \bibinfo{journal}{\bibinfo{title}{Effective public health measures to mitigate the spread of covid-19: a systematic review}}.
\newblock {\emph{\JournalTitle{BMC Public Health}}} \textbf{\bibinfo{volume}{21}}, \bibinfo{pages}{1015}, \doiprefix\url{10.1186/s12889-021-11111-1} (\bibinfo{year}{2021}).

\bibitem{Konstantinos}
\bibinfo{author}{Gkiotsalitis, K.} \& \bibinfo{author}{Cats, O.}
\newblock \bibinfo{journal}{\bibinfo{title}{Public transport planning adaption under the covid-19 pandemic crisis: literature review of research needs and directions}}.
\newblock {\emph{\JournalTitle{Transport Reviews}}} \textbf{\bibinfo{volume}{41}}, \bibinfo{pages}{374--392}, \doiprefix\url{10.1080/01441647.2020.1857886} (\bibinfo{year}{2021}).
\newblock \eprint{https://doi.org/10.1080/01441647.2020.1857886}.

\bibitem{bota2017b_modeling}
\bibinfo{author}{Bóta, A.}, \bibinfo{author}{Gardner, L.} \& \bibinfo{author}{Khani, A.}
\newblock \bibinfo{title}{Modeling the spread of infection in public transit networks: a decision-support tool for outbreak planning and control}.
\newblock In \emph{\bibinfo{booktitle}{Transportation Research Board 96th Annual Meeting}} (\bibinfo{year}{2017}).

\bibitem{hajdu2020discovering}
\bibinfo{author}{Hajdu, L.}, \bibinfo{author}{Bóta, A.}, \bibinfo{author}{Krész, M.} \emph{et~al.}
\newblock \bibinfo{journal}{\bibinfo{title}{Discovering the hidden community structure of public transportation networks}}.
\newblock {\emph{\JournalTitle{Networks and Spatial Economics}}} \textbf{\bibinfo{volume}{20}}, \bibinfo{pages}{209--231}, \doiprefix\url{10.1007/s11067-019-09476-3} (\bibinfo{year}{2020}).

\bibitem{Khani2013}
\bibinfo{author}{Khani, A.}
\newblock \emph{\bibinfo{title}{Models and solution algorithms for transit and intermodal passenger assignment (development of fast-trips model)}}.
\newblock \bibinfo{type}{Phd dissertation}, \bibinfo{school}{University of Arizona}, \bibinfo{address}{Tucson, AZ, USA} (\bibinfo{year}{2013}).

\bibitem{Khani2014}
\bibinfo{author}{Khani, A.}, \bibinfo{author}{Beduhn, T.}, \bibinfo{author}{Duthie, J.}, \bibinfo{author}{Boyles, S.} \& \bibinfo{author}{Jafari, E.}
\newblock \bibinfo{title}{A transit route choice model for application in dynamic transit assignment}.
\newblock In \emph{\bibinfo{booktitle}{Innovations in Travel Modeling}} (\bibinfo{address}{Baltimore, MD}, \bibinfo{year}{2014}).

\bibitem{GTFS2024}
\bibinfo{author}{{General Transit Feed Specification}}.
\newblock \bibinfo{title}{General transit feed specification (gtfs)} (\bibinfo{year}{2024}).
\newblock \bibinfo{note}{Accessed: 2024-09-19}.

\bibitem{SFCTA2024}
\bibinfo{author}{{San Francisco County Transportation Authority}}.
\newblock \bibinfo{title}{Sf-champ modeling} (\bibinfo{year}{2024}).
\newblock \bibinfo{note}{Accessed: 2024-09-19}.

\bibitem{park2021risk}
\bibinfo{author}{Park, J.} \& \bibinfo{author}{Kim, G.}
\newblock \bibinfo{journal}{\bibinfo{title}{Risk of covid-19 infection in public transportation: The development of a model}}.
\newblock {\emph{\JournalTitle{Int. J. Environ. Res. Public Health}}} \textbf{\bibinfo{volume}{18}}, \bibinfo{pages}{12790}, \doiprefix\url{10.3390/ijerph182312790} (\bibinfo{year}{2021}).

\bibitem{chen2021}
\bibinfo{author}{Chen, L.} \emph{et~al.}
\newblock \bibinfo{journal}{\bibinfo{title}{Estimation of the sars-cov-2 transmission probability in confined traffic space and evaluation of the mitigation strategies}}.
\newblock {\emph{\JournalTitle{Environmental Science and Pollution Research}}} \textbf{\bibinfo{volume}{28}}, \bibinfo{pages}{42204--42216}, \doiprefix\url{10.1007/s11356-021-13707-w} (\bibinfo{year}{2021}).

\end{thebibliography}

\end{document}